\documentstyle[12pt]{article}
\begin{document}

\pagestyle{plain}
\thispagestyle{empty}
\vspace*{-10mm}\hskip
\begin{flushright}
 BROWN-HET-931\\
TA-505
\end{flushright}

\vspace{2.5cm}
\begin{center}
{\Large \bf LEP Electroweak Physics\footnote{Presented at the 14th
International Workshop on Weak Interactions and Neutrinos, 19-24 July
1993, Seoul National University, Seoul, Korea, and supported in part by the
US Department of Energy Contract DE-FG02-91ER40688-Task A} } \\
\vglue 10mm
{\bf Kyungsik Kang } \\
\vglue 5mm
{\it Department of Physics, Brown University\footnote{Permanent address}
, Providence, RI, USA, }\\
\vglue 1mm
{\it Center for Theoretical Physics, Seoul National University }\\
\vglue 1mm
{and} \\
\vglue 1mm
{\it Center for High Energy Physics, Korea Advanced Institute of }\\
\vglue 1mm
{\it Science and Technology. }\\
\vglue 20mm
{\bf ABSTRACT} \\
\vglue 10mm
\begin{minipage}{14cm}
{\normalsize
This is an overview of the electroweak physics achieved at LEP.
After a brief review of the standard model and the calculation
of the radiative corrections both in the $\overline{MS}$ and
on-shell schemes, we introduce the $Z-$ decay parameters
around the resonance energy and discuss the status of the
standard model fits for them, along
with some selected highlights of the measurements as well as new
particle searches.
}
\end{minipage}
\end{center}
\vfill
\newpage
\setcounter{page}{1}
\begin{center}
{\bf LEP Electroweak Physics } \\
\vglue 8mm
{\bf Kyungsik Kang } \\
\vglue 1mm
{\it Department of Physics, Brown University, Providence, RI, USA, }\\
\vglue 1mm
{\it Center for Theoretical Physics, Seoul National University }\\
\vglue 1mm
{and} \\
\vglue 1mm
{\it Center for High Energy Physics, Korea Advanced Institute of }\\
\vglue 1mm
{\it Science and Technology. }\\
\vglue 8mm
{\bf ABSTRACT} \\
\vglue 4mm
\begin{minipage}{14cm}
{\small
This is an overview of the electroweak physics achieved at LEP.
After a brief review of the standard model and the calculation
of the radiative corrections both in the $\overline{MS}$ and
on-shell schemes, we introduce the $Z-$ decay parameters
around the resonance energy and discuss the status of the
standard model fits for them, along
with some selected highlights of the measurements as well as new
particle searches.
}
\end{minipage}
\end{center}
\vglue 10mm
\baselineskip20pt

%
\section{Introduction}
Recent high precision experiments [1] on $Z^0$ decays by the LEP groups allow
us to
study precise tests of the standard model and compare to the theoretical
expectations.
During $1990 - 1992$, each of the four LEP experiments has accumulated more
than
1.5 million $Z^0$ decays corresponding to a  total integrated luminosity of
about 45 $pb^{-1}$.
In 1991, the method of resonant depolarization was used to calibrate the beam
energy by measuring the frequency with which the spins of transversely
polarized electrons process about the vertical axis of the bending field.
Combined with a better understanding of the properties of the LEP magnets
and RF system, a precise absolute energy scale of the LEP beams was achieved
and a common systematic error of the four experiments to the LEP energy
uncertainty is now 6.3 MeV on $M_Z$ compared to 20 MeV in 1990.
As a consequence, $M_Z$ is now known to an unprecedent precision of better than
$0.01\%$, i.e., $M_Z=91.187\pm 0.007 $ GeV, which is so precise that
even the tidal effect of the moon and sun, which cause the LEP ring size to
change by a few parts in $10^8$ and thus cause the change in energy by about
$8 $ MeV, has to be corrected [2].

There are a number of good review reports [3] on the LEP experiments from which
I will freely quote the numbers.
After a brief review of the standard model related to the $Z$-decay parameters
the talk will start with the electroweak radiative corrections which must be
taken into the account to compare with the LEP experiments.
Both the on-shell and modified minimal subtraction renormalization scheme
will be discussed.
We then will compare the $Z$-resonance and decay parameters with the fits to
the standard model.
Particular emphasis will be given to the number of neutrino species, the
evolution of the QCD coupling constant $\alpha_s$, the $B$-decays from $Z^0$
and new particle searches at LEP.
\section{Brief Review of the Standard Model and Radiative Corrections}
The standard model for elementary particle interactions is based on the gauge
group $SU(3)\times SU(2)\times U(1)$ for three generations of leptons and
color quarks.
The model is developed to incorporate all confirmed symmetries of the
elementary particle interactions, the success of the quark model, QED, and
charged-current weak interactions.
Its success rests on the correct predictions of the neutral current weak
interactions and the properties of the gauge bosons
of gluons, $W^{\pm }$ and $Z$ besides the photon.
Indeed the $e^{+}e^{-}$ collider LEP is designed to study the final states that
are produced by the neutral gauge boson $Z^0$ and photon at the intermediate
stage.

In the standard model, the basic interaction Lagrangian is given by [4]
\begin{eqnarray}
L_{int} = &-& \frac{g_2}{\sqrt{2}}\sum_{i}\bar{\psi}_{i}\gamma^{\mu}
\frac{(1+\gamma_5)}{2}(T_{+}W_{\mu}^{-}+T_{-}W_{\mu}^{+})\psi_{i} \nonumber\\
 &-& e\sum_{i}Q_{i}\bar{\psi}_{i}\gamma^{\mu}\psi_{i}A_{\mu} \nonumber\\
&-& \frac{g_2}{\cos {\theta_W}}Z_{\mu}\sum_{i}\bar{\psi}_{i}\gamma^{\mu}
(T_3-Q_{i}\sin^2{\theta_W}+\gamma_5T_3)\psi_{i} \nonumber\\
&-& \frac{g_2}{2}\sum_{i}\frac{m_i}{M_W}\bar{f_i}hf_{i}
\end{eqnarray}
where $\psi_{i}$ represents a left-handed doublet of quarks and leptons, $h$
is the real part deviation from the Higgs  potential minimum of the neutral
component of a scalar doublet ( there is just one doublet in the minimal
standard model), and $f_i$ stands for a fermion field of quarks and leptons
with mass $m_i$.
Also $T_i$ is the weak isospin matrix $\tau_i/2$, $Q_i$ is the charge matrix
and the (unrenormalized) $SU(2)$ coupling constant $g_2$ is given by $e/\sin
{\theta _W}$ where $\theta_W$ is the mixing angle of $SU(2)$ neutral gauge
field and $U(1)$ so as to make $Z_{\mu }$ and $A_{\mu }$ to be the correct
mass eigenstate of the neutral gauge boson and photon respectively.
In particular, the photon coupling to charged fermions ia given by the familiar
QED current, $ieQ_f\gamma ^{\mu}$, in terms of the charge of the fermion $eQ_f$
and the neutral gauge boson coupling to a fermion $f$ is given by
\begin{equation}
L_{Zf\bar{f}} = -\frac{ie}{\sin^2\theta_W}\bar{f}\gamma^{\mu}
(T_3^f-2Q_f\sin^2{\theta_W}+\gamma^5T_3^{f})f
\end{equation}
which gives, in particular, the vector and axial-vector couplings of the
electron to be $g^e_V = 2\sin^2{\theta_W}-\frac{1}{2}, g_A^e= -\frac{1}{2}.$
The mass of the weak gauge bosons follows from the spontaneous $SU(2)\times
U(1)$ gauge symmetry breaking and Higgs mechanism and is related to each other
by $M_W =M_Z\cos {\theta_W}.$ By comparing the low-momentum transfer limit of
the charged-current interaction mediated by $W$ exchanges to the traditional
four-point Fermi interactions, one obtains the relationship $g_2^2/8M_W^2
=G_F/\sqrt{2}$ so that the tree-level approximation
\begin{equation}
\sin^2{\theta _W}=1-M_W^2/M_Z^2 = \pi \alpha/\sqrt{2}G_FM ^2_W
\end{equation}
holds where $\alpha =e^2/4\pi$.
In principle, this relationship should be modified by renormalized parameters
through radiative corrections which can be written as the on-shell relation [5]
\begin{equation}
\sin^2{\theta_W} = \frac{\pi \alpha }{\sqrt{2}G_FM_W^2(1-\Delta r)}
\end{equation}

The Fermi coupling constant $G_F$ is determined from the $\mu $-lifetime
$\tau _{\mu}$ including the QED corrections,
\begin{equation}
\frac{1}{\tau _{\mu}}=\Gamma (\mu \rightarrow e\nu_{\mu}\bar{\nu }_e(\gamma))
=\frac{G_F^2m_{\mu }^5}{192\pi^3}f(\frac{m_e^2}{m_{\mu }^2})
\left[ 1+\frac{\alpha}{2\pi}(\frac{25}{4}-\pi^2)(1+\frac{2\alpha}{3\pi}\ln
{\frac{m_{\mu}}{m_e}})\right]
\end{equation}
where $f(x)$ is the three-body phase space factor
\begin{eqnarray*}
f(x)=1-8x+8x^2-x^4-12x^2\ln {x}
\end{eqnarray*}
upon using $\alpha^{-1} = 137.0359895$ as deduced from the classical Thomson
cross section $\sigma = e^4/6m_e^2$ in $k^2=0, k^0\rightarrow 0$ limit of the
QED result or from the electron magnetic moment anomaly $(g-2)_{exp}$, and
the measured muon lifetime $\tau _{\mu }=(2.197035\pm 0.000040)\times 10^{-6}
sec $.
One obtains $G_F=1.16639\times 10^{-5}  \mbox{GeV}^{-2}$ so that $\pi \alpha
/\sqrt{2}G_F = (37.28 \mbox{GeV})^2$.
If $M_W$ is known to the same accuracy as $M_Z$, the needed amount of radiative
corrections $\Delta r$ can be determined from (4) with the on-shell value of
$\sin ^2{\theta _W}$ as given by $1-M_W^2/M_Z^2$.
But the uncertainty in $M_W$ can cause as much as $10\%$ variation in the
on-shell $\sin^2{\theta_W}$, i.e., $\sin^2{\theta_W}=0.2160$ for
$M_W = 80.22 $ GeV while $\sin^2{\theta _W}=0.2321$ for $M_W=79.91$ GeV, and
this
makes the estimate of $\Delta r$ to be sensitive to the value of $M_W$.
Theoretically $\Delta r$ can be calculated from $G_F/\sqrt{2}$ obtained in the
$k^2=0$ limit of $\mu \rightarrow e\bar{\nu _e}\nu _{\mu }$ with radiative
corrections.
This procedure calls for the renormalized couplings and masses which gives in
the one-loop order [6]
\begin{eqnarray}
\frac{G_F}{\sqrt{2}}&=&\frac{\pi \alpha }{2M_W^2\sin^2{\theta _W}}
\left[1+2\frac{\delta e}{e}-\cot ^2{\theta _W}\left(\frac{\delta
M_Z^2}{M_Z^2}-\frac{\delta M_W^2}{M_W^2}\right) +\frac{\Sigma _{WW}(0)-\delta
M_W^2}{M_W^2}\right. \nonumber \\
& & +\left. (\Delta r)_{vertex}+(\Delta r)_{box}\right]
\end{eqnarray}
where the various counter terms $\frac{\delta e}{e},\frac{\delta M^2}{M^2}$
etc. are defined as
\begin{eqnarray}
e^2\rightarrow e^2\left(1+2\frac{\delta e}{e}\right),~~~~M^2\rightarrow M^2
\left(1+\frac{\delta M^2}{M^2}\right) \nonumber \\
\sin^2{\theta_W}\rightarrow \sin^2{\theta_W}
\left[1+\cot^2{\theta_W}\left(\frac{\delta M_Z^2}{M_Z^2}-\frac{\delta
M_W^2}{M^2_W}\right)\right]
\end{eqnarray}
which are to be fixed by the renormalization scheme adopted.
Here $(\Delta r)_{vertex}$ and $(\Delta r)_{box}$ are the vertex and the box
diagram corrections and $\Sigma _{WW}$ is the transverse part of the $W$
self-energy.
Thus the quantity $\Delta r$ is the finite combination of loop diagrams and
counter terms.

Actual calculation of the loop insertion diagrams requires a regularization
scheme [7] in which the dimension 4 is replaced by a lower dimension $D$ such
that the integrals are convergent
\begin{equation}
\int \frac{d^4k}{(2\pi )^4}\cdot \cdot \cdot \rightarrow
\mu ^{4-D}\int \frac{d^Dk}{(2\pi )^D}\cdot \cdot \cdot
\end{equation}
and then the limit $D\rightarrow 4$ is to be taken at the end whereby
identifying and subtracting the divergent terms
$\Delta = \frac{1}{D-4} + \frac{\gamma}{2}-\ln {\sqrt{4\pi }}$ for $\mu >m_f$
and $\Delta + \ln {m_f/\mu }$ for $\mu < m_f$ in such a way that a physical
quantity can be matched smoothly at $\mu = m_f$.
In (8), $\mu ^{4-D}$ is to insure the coupling constant in front of the
integral to be kept as dimensionless.

Turnning to the counter terms (6) and (7), firstly we want to adopt the charge
renormalization condition in the full electroweak theory so as to reproduce the
classical charge in the Thomson cross section in the limit $k^2\rightarrow 0$
and $k^0\rightarrow 0$.
This gives
\begin{equation}
\frac{\delta e}{e} =\frac{1}{2}\Pi ^{\gamma }(0)-\tan {\theta _W}
\frac{\Sigma ^{\gamma Z}(0)}{M^2_Z}
\end{equation}
where the first term is the vacuum polarization of the photon related to the
diagonal photon self-energy by $\Sigma ^{\gamma \gamma}(q^2)=q^2\Pi ^{\gamma}
(q^2)$ and the second term is due to the $\gamma-Z$ mixing propagator $\Sigma
^{\gamma Z}$.
The mass renormalization conditions to insure that $M_{W,Z}$ are the physical
$Z$ or $W$ masses fix the counter terms
\begin{equation}
\delta M^2_{W,Z} = Re \Sigma ^{W,Z}(M^2_{W,Z})
\end{equation}
The fact that the vacuum polarization corrects the electric charge allows us to
define the concept of the "running charge".
This concept is however scheme dependent, i.e., in the on-shell scheme,
\begin{equation}
\alpha (M_Z^2) = \frac{\alpha }{1+Re \hat{\Pi }^{\gamma }(M_Z^2)}
\end{equation}
where $\hat{\Pi}^{\gamma }(q^2)=\Pi ^{\gamma}(q^2)-\Pi ^{\gamma }(0)$, the
renormalized vacuum polarization, and in the $\overline{MS}$ scheme [8],
\begin{equation}
\hat{\alpha}(M_Z^2) = \alpha/[1-\Pi^{\gamma}(0)
+2\tan {\theta_W}\Sigma ^{\gamma Z}(0)/M^2_Z]_{\overline{MS}}
\end{equation}
in which $\Pi^{\gamma }(0)_{\overline{MS}}$ is splitted in terms of the
renormalized quantity $\hat{\Pi}^{\gamma}$ as
\begin{equation}
\Pi^{\gamma}(0)_{\overline{MS}} = -Re\hat{\Pi^{\gamma}}(M^2_Z)
+Re \Pi^{\gamma}(M^2_Z)
\end{equation}
with the finite part $Re \hat {\Pi }^{\gamma}(\mu ^2)$ given by
\begin{equation}
Re \hat{\Pi}^{\gamma}(\mu ^2)= -\frac{\alpha}{3\pi}
\left \{\sum_{f}Q_f^2\ln {(\frac{\mu }{m_f})^2}+\frac{9}{4}\ln
{(\frac{M_W}{\mu})^2}
-\frac{1}{2}\right \}
\end{equation}
and $\tan {\theta _W}\Sigma ^{\gamma Z}(0)/M_Z^2 = -\frac{\alpha }{\pi}
\ln {(M_W/\mu )^2}$.
Thus if we set
\begin{equation}
\alpha (M^2_Z) = \alpha /(1-\Delta \alpha )
\end{equation}
we get in the $\overline{MS}$ scheme
\begin{equation}
\Delta \alpha _{\overline{MS}}=\frac{2\alpha}{3\pi}
\left \{\sum_{f}Q_f^2\ln {(\frac{M_Z}{m_f})}+\frac{21}{4}\ln {\cos {\theta_W}}
-\frac{1}{4}\right \}
\end{equation}
while in the on-shell scheme
\begin{equation}
\Delta \alpha (M_Z^2)=\frac{2\alpha}{3\pi}
\left \{\sum_{f}Q_f^2\left(\ln {(\frac{M_Z}{m_f})}-\frac{5}{6}\right) \right \}
\end{equation}
The leptonic contributions in (16) and (17) are easy to calculate but
due to the ambiguities in the quark masses the hadronic contributions can
better be estimated from the experimental data of $R(s)=\sigma
_{tot}(e^{+}e^{-}
\rightarrow hadrons)/\sigma (e^{+}e^{-}\rightarrow \mu ^{+}\mu ^{-})$
up to $40 \mbox{GeV}$ with the help of a dispersion relation
\begin{equation}
\hat{\Pi}^{\gamma}_{had}(M_Z^2)=\frac{\alpha }{3\pi}M_Z^2
\int ^{\infty}_{4m_{\pi}^2} ds \frac{R(s)}{s(s-M_Z^2)}
\end{equation}
which gives $Re\hat{\Pi}^{\gamma}_{had}(M_Z^2)= -0.0288(9)$.
Combining this with the leptonic part, one obtains $\Delta \alpha (M_Z^2)=
0.0602(9)$ so that $\alpha (M^2_Z)=(128.78\pm 0.12)^{-1}$
whereas $\Delta \alpha _{\overline{MS}}=0.0677(9)$ so that
$\hat{\alpha }(M_Z^2)=(127.76\pm 0.12)^{-1}$.
Note that $\hat{\alpha }(M_Z^2)$ in (12) contains the $W$ contributions in the
$\overline{MS}$ renormalization, i.e., after subtracting the $D-4$ poles and
associated constants in dimensional regularization, and it has a $0.8 \%$
difference from the conventional QED renormalization in the on-shell scheme
$\alpha (M_Z^2)$.

The radiative correction $\Delta r$ in (6) can be cast into
\begin{equation}
\Delta r = \Delta \alpha -\cot^2{\theta_W}\Delta \rho + (\Delta r)_{rem}
\end{equation}
where $\Delta \rho $ is the combination $\Sigma ^{ZZ}(0)/M_Z^2-\Sigma ^{WW}(0)
/M_W^2$ entering the NC/CC neutrino cross section ratio.
The main contribution to $\Delta \rho $ comes from the heavy t-quark
\begin{equation}
\Delta \rho = \frac{3\alpha }{16\pi \sin^2{\theta _W}}\frac{m_t^2}{M_W^2}
=\frac{3\alpha }{16\pi \sin^2{\theta _W}\cos^2{\theta _W}}
\left(\frac{m_t}{M_Z}\right)^2
\end{equation}
while the Higgs boson contribution is a part of the remainder
\begin{equation}
(\Delta r)^{Higgs}_{rem} = \frac{\alpha }{16\pi \sin ^2\theta _W}\frac{11}{3}
    \left(\ln {\frac{M^2_H}{M^2_W}}-\frac{5}{6}\right)
\end{equation}
along with a logarithmic term in $m_t$
\begin{equation}
(\Delta r)^t_{rem} = -\frac{\alpha}{4\pi \sin^2\theta_W}
\left(\cot^2\theta_W-\frac{1}{3}\right )\ln {\frac{m_t}{M_Z}}
\end{equation}
coming from $t\bar{t}W$ and $WWt$ vertex corrections with one-loop order.
For $m_t$ around ($1\sim 2) ~~M_W$, (20) and (21) are the main contributions
to $\Delta r$ besides the QED correction $\Delta \alpha $.
There is also additional light fermion contribution term in the remaining
part
\begin{equation}
(\Delta r)^{l.f.}_{rem} = \frac{\alpha}{4\pi \sin^2\theta_W}
\left(1-\cot^2\theta_W\right )\ln {\cos {\theta_W}}
\end{equation}
besides that in $\Delta \alpha $.
For $M_W$ of about $80 $ GeV, $\Delta r$ is of the size $0.04 - 0.07$ from
(4) and therefore $(\Delta r)_{QED}=\Delta \alpha (M_Z^2)\approx 0.06$ is a
major component of the needed amount of the  radiative correction.
\section{Cross Section for $e^{+}e^{-}\rightarrow f\bar{f}$ and $Z-$ decay
Parameters}
The differential cross section for $e^{+}e^{-}\rightarrow f\bar{f}$ through
$\gamma $ and $Z$ can be calculated from (1) in the standard model, which
contains $\gamma -Z$ interference term in addition to $\gamma-$ and $Z-$
pole terms [9].
To a very good approximation the cross section for $e^{+}e^{-}\rightarrow
f\bar{f}$ after one-loop electroweak radiative corrections can be obtained from
the tree-level results.
Namely, starting with the partial cross section
\begin{eqnarray}
\sigma_f(s) &=& \sigma^{P}_f\frac{s\Gamma^2}
                             {(s-M_Z^2)^2+s^2\Gamma^2/M_Z^2}
+\frac{4\pi \alpha^2}{3s}N_c^fQ^2_f \nonumber \\
&-& \frac{2\sqrt{2}\alpha }{3}G_{F}Q_fN^f_cv_ev_f
\frac{(s-M_Z^2)M_Z^2}{(s-M^2_Z)^2+s^2\Gamma^2/M_Z^2}
\end{eqnarray}
where
\begin{eqnarray}
\sigma^P_f &=& \frac{12\pi \Gamma_e\Gamma_f}
                 {M_Z^2\Gamma(1+\frac{3\alpha}{4\pi})} ,\\
 \Gamma &=& \sum_f\Gamma (Z\rightarrow f\bar{f})=\sum_f\Gamma_f
\end{eqnarray}
and
\begin{equation}
\Gamma_f = \frac{G_{F}M_Z^3}{24\sqrt{2}\pi}\beta_fR^f_cR_QN_c^f
\left\{v_f^2(1+2m_f^2/M_Z^2)+a^2_f(1-4m_f^2/M_Z^2)\right\},
\end{equation}
one makes the following replacements
\begin{eqnarray}
\alpha &\rightarrow &\alpha (M_Z^2) ~~;~~ (16\pi
\sin^2\theta_W\cos^2\theta_W)^{-1}
\rightarrow \sqrt{2}G_{F}M_Z^2\rho /16\pi\alpha (M_Z^2) ; \nonumber \\
a_f&\rightarrow &\sqrt{\rho}~2~T_3^f ~~;~~ v_f\rightarrow \sqrt{\rho}~2~
\left(T_3^f-2Q_f\sin^2\theta_W^{eff}\right); \nonumber \\
\sin^2\theta_W &\rightarrow &\sin^2\theta^{eff}_W =\frac{1}{2}
\left(1-\sqrt{1-4\pi \alpha(M_Z^2)/\sqrt{2}G_{F}M_Z^2\rho }\right)
\end{eqnarray}
where
\begin{equation}
\rho = 1+\frac{3G_{F}m_t^2}{8\sqrt{2}\pi ^2}-\frac{11G_{F}M_Z^2\sin^2
\theta_W^{eff}}{12\sqrt{2}\pi^2}\left(\ln {\frac{m_H}{M_W}}-\frac{5}{12}\right)
\end{equation}
In (28), $\beta_f = (1-4m_f^2/M^2_Z)^{1/2}, R^f_Q = 1+\frac{3\alpha }{4\pi}
Q^2_f, R_c=1+\alpha_s/\pi +1.409(\alpha_s/\pi)^2-12.805(\alpha_s/\pi)^3$
for quarks and $R_c=1$ for leptons, and the vector and axial vector couplings
are given by
\begin{eqnarray}
v_f&=&2g_V^f=2T^f_3, \nonumber \\
a_f&=&2g^f_A = 2\left(T^f_3-2Q_f\sin^2\theta_W\right)
\end{eqnarray}
The tree-level results are calculable in terms of $\alpha, G_{F}, M_Z$ from
(24)-(27) but the one-loop terms are of the order $\alpha/\pi=0.0023$
while the typical experimental precision at LEP is a few $10^{-3}$.
Thus one needs to dress the Born $\gamma $ and $Z$ exchange propagators and
to modify the $\gamma f\bar{f}$ and $Zf\bar{f}$ vertices by one loop diagrams.
The one-loop box diagrams involving $\gamma \gamma,\gamma Z, ZZ$ and $WW$
exchanges for $e^{+}e^{-}\rightarrow f\bar{f}$ should also be included for
consistency. The infrared divergence coming from massless virtual photon
loop attached to the external charged fermion in the gauge invariant subset
of one-loop QED graphs is taken care of by adding the real photon
bremsstrahlung
diagram. The non-QED or weak correction diagrams are free of the infrared
divergence. The ultraviolet divergences are removed by the counter terms as a
consequence of renormalizability as discussed in the previous section.
Because of the factorization of the weak corrections at the $Z$ pole,
it can be cast into an overall renormalization of the
$Zf\bar{f}$ vertex and a renormalization of $\sin^2\theta_W$ in the neutral
vector coupling [10],
\begin{eqnarray}
G_{F}&\rightarrow &\rho G_{F} \nonumber \\
\sin^2\theta_W &\rightarrow &\kappa_f\sin^2\theta_W^{eff}
\end{eqnarray}
where
\begin{equation}
\kappa_f=1+\cot^2\theta_W\Delta \rho +\Delta \kappa_{f,vertex}+
\Delta \kappa_{f,rem}
\end{equation}
the leading part of
$\Delta \rho $ being given by $\rho -1$ from (20).
In addition there is a subleading term from the self-energy terms
\begin{eqnarray}
(\Delta \kappa )_{t,rem} = \frac{\sqrt{2}G_FM_W^2}{4\pi^2}
\left\{ \frac{1}{3} (\frac{3}{\sin^2\theta_W}-2) \ln {(\frac{m_t}{M_W})}
\right\}
 \\
(\Delta \kappa )_{Higgs,rem} = \frac{\sqrt{2}G_FM_W^2}{16\pi^2}
\left\{ -\frac{10}{3}(2\ln {(\frac{m_H}{M_W})}-\frac{5}{6})\right\}
\end{eqnarray}

$\Delta \kappa_{f,vertex}$ is negligible for $f\neq b$ but due to $Wb\bar{t}$
couplings the $Zb\bar{b}$ vertex gets a heavy t-quark contribution,
\begin{equation}
\Delta \kappa _{b,vertex}=\frac{\sqrt{2}G_{F}M_W^2}{16\pi^2}
\left\{2(m_t/M_W)^2+\frac{1}{3}\left(16+\frac{1}{\cos^2\theta_W}\right)\right.
\left. \ln{\frac{m_t^2}{M_W^2}}+\cdot\cdot\cdot \right \}
\end{equation}
Because of these extra contributions from $t\bar{t}W$ and $W^{+}W^{-}t$
triangle graphs, $\Gamma_{b\bar{b}}$ is different from $\Gamma_{d\bar{d}}$
or $\Gamma_{s\bar{s}}$ by about $2\%$ for $m_t$ around $150 $ GeV
and by more for a heavier $m_t$.
The precise measurements of $\Gamma_{b\bar{b}}$ can provide further test
of the standard electroweak theory.
Besides $\Gamma_Z,~\Gamma_{l\bar{l}},~\Gamma_{had},~\sigma_f^{P}$ and the
number of light neutrino specise $N_{\nu}\equiv \Gamma_{inv}/\Gamma_{\nu
\bar{\nu}}$, a number of asymmetry parameters such as the forward-backward
asymmetry $A_{FB}(f)$, for $e^{+}e^{-}\rightarrow f\bar{f}, \tau -$
polarization $P_{\tau}$ and the polarized F-B asymmetry
 $A^{pol}_{FB}(\tau) $
are measured at LEP.
$A_{FB}(f)$ are obtained by fits of the angular distributions
\begin{equation}
\frac{d\sigma}{d\cos {\theta}} \propto 1+\cos^2 {\theta}+
\frac{8}{3}A_{FB}(f)\cos {\theta}
\end{equation}
where in the Born level
\begin{eqnarray*}
A_{FB}(f) &=& \frac{3}{4}A_eA_f \\
\mbox{with}~~~~~~~ A_f &=& \frac{2g_V^fg_A^f}{(g_V^f)^2+(g_A^f)^2} = -P_f
\end{eqnarray*}
In particular, from $e^{+}e^{-}\rightarrow \tau^{+}\tau^{-}, P_{\tau}=
-A_{\tau}$ and $A^{pol}_{FB} = -\frac{3}{4}A_e$.

There are a number of works that fitted the cross-sections and $A_{FB}(l)$
simultaneously by using the programs like ZFITTER [11] within the framework
of the standard electroweak model.
They are all in good agreement with each other within the accuracy of the
measurements.
The disrepancy in global fits from the predictions based on $\alpha , G_F,
M_Z,$
and the measured partial and total widths signal the need of new physics beyond
the standard electroweak model.
But as of now, they can provide only the constraints on t-quark and Higgs
scalar
masses.
According to the results of the standard model fit by Monica Pepe Altarelli [3]
using the programs MIZA [12] and ZFITTER to the LEP data of $\Gamma ,
\sigma^{P}_{had}, R_{l}=\Gamma_{had}/\Gamma_{l\bar{l}}$ and $A_{FB}(l)~~
(l=e,\mu ,\tau),$
the LEP data supports the standard model remarkable way with $m_t=
163^{+19+18}_{-21-20}$ for $m_H=60 - 1,000 $ GeV and
$\alpha_s = 0.127\pm 0.007\pm 0.002$ with $\chi^2/d.o.f =4.6/7$.
Here the first error is statistical and second is the variation due to
$m_H$ between 60 and $1,000 $ GeV.
The $W$ mass determined is $M_W = 80.23\pm 0.12\pm 0.02 $ GeV.
Inclusion of the data obtained outside the LEP groups gives similar
result, $m_t=160^{+16+18}_{-18-19} \mbox{GeV}, \alpha_s =0.127\pm 0.007\pm
0.002 $ with $\chi ^2/d.o.f. = 8.5/14$.
In particular, the upper and lower limits at $95\% CL, 203 \mbox{GeV} >
m_t > 109 \mbox{GeV},$ are obtained from $m_H=1,000 $ GeV and  60 GeV
respectively.
Other fits give more or less the same result, for example, Langacker [3]
gives $m_t = 150^{+19}_{-24} $ GeV for $m_H = 60 - 1,000 $ GeV and $\alpha_s
=0.12\pm 0.01.$

The agreement of the LEP data with the standard model fit is impressive from
the determined values of the $Z-$ decay parameters at the resonance energy $M_Z
=91.187\pm 0.007 $ GeV.
In particular, width $\Gamma _{inv}=\Gamma -\sum \Gamma _l - \Gamma _{had}
=N_{\nu}\Gamma_{\nu}$ is $ 2.99\pm 0.03$ to be compared to $N_{\nu }=3$.
The invisible width can be written in terms of the width ratios
\begin{equation}
\frac{\Gamma_{inv}}{\Gamma_{l}} = \sqrt{\frac{12\pi R_l}{M^2_Z\sigma^P_{had}(1+
\frac{3}{4}\frac{\alpha }{\pi })}}-3-R_l = N_{\nu }\frac{\Gamma_{\nu}}
{\Gamma_{l}}
\end {equation}
so that $N_{\nu }=(\Gamma _{inv}/\Gamma_l)/(\Gamma_{\nu}/\Gamma_l)$ in which
the standard model value for $(\Gamma_{\nu}/\Gamma_l)=1.993\pm 0.004$
and measured value $R_l=20.83\pm 0.06$ are used.
Note that $\Gamma_{\nu}/\Gamma_l$ is almost independent of the radiative
corrections.
Also the axial-vector coupling of the lepton is measured to be
$g^l_A = -0.4999\pm 0.0009$ compared to $g_A^l=T_3^3=-\frac{1}{2}$.
The effective electroweak mixing parameter $\sin ^2\theta_W^{eff}$ is defined
from the ratio of the vector and axial-vector couplings
\begin{equation}
g^l_V/g_A^l = 1-4\sin^2\theta_W^{eff}
\end{equation}
in the standard model and this can be extracted from the various asymmetry
at LEP to be
\begin{eqnarray*}
\sin ^2{\theta^{eff}_W}=0.2319\pm 0.0007
\end{eqnarray*}
The overall goodness of the standard model fit to the LEP data is
impressive and shows no significant deviation from the standard electroweak
model and therefore leaving no room for new physics.
\section{Other Measurements at LEP}
(a) Single photon events originating from the initial state radiation in
$e^{+}e^{-} \rightarrow \nu \bar{\nu }\gamma $ are used by L3 to determine the
number of light neutrino species, i.e., $N_{\nu }=3.14\pm 0.024(\mbox{stat.})
\pm 0.012$, in very good agreement with $N_{\nu }=3$, \\
(b) $\alpha_s(M_Z) $ can be determined from the ratio of the hadron width to
the leptonic width measured at LEP,
$R_l=\Gamma_{had}/\Gamma_l = R_0 R_c$, where $R_c$ is the QCD correction
calculated
to $ O(\alpha^3)$ and $R_0 = 19.95\pm 0.03$ from the standard model.
This gives $\alpha_s(M_Z) $ from all OPAL measurements is
$\alpha_s(M_Z) =1.22^{+0.006}_{-0.005}$, while from the study of the scaling
violations in fragmentation functions measured at LEP and at lower
$e^{+}e^{-}$ energies, DELPHI gives $\alpha_s(M_Z) = 0.119\pm 0.006$. \\
(c) The evolution of the couplings in the standard model is dictated by the
running slopes $b_i$ determined by the loops present in the boson propagator as
\begin{equation}
\mu \frac{\partial \alpha_i(\mu )}{\partial \mu }=b_i
\frac{\alpha^2_i(\mu ) }{2\pi } +\cdot \cdot \cdot
\end{equation}
so that to the lowest approximation the inverse coupling $\alpha _i^{-1}$
has a linear slope with the logarithm of the energy
\begin{equation}
\alpha ^{-1}_i (\mu )=\alpha ^{-1}_i(\mu _0)-\frac{b_i}{2\pi }
\ln {\mu /\mu_0}
\end{equation}
Here $\alpha_1 =\frac{5}{3}\frac{\alpha}{\cos^2{\theta_W}}, \alpha_2=\alpha
/\sin^2{\theta_W}, \alpha_3=\alpha_s$ and
\begin{eqnarray}
 \left(\begin{array}{c}
 b_1 \\
 b_2 \\
 b_3
\end{array} \right)  =
 \left(\begin{array}{c}
 0 \\
 -22/3 \\
 -11
\end{array} \right) +n_{fam}
\left(\begin{array}{c}
 4/3 \\
 4/3 \\
 4/3
\end{array} \right) +n_{Higgsdoublets}
\left(\begin{array}{c}
 1/10 \\
 1/6 \\
 0
\end{array} \right)
\end{eqnarray}
in the standard model for $\mu > m_t$ with $n_{fam}=3 $ and $n_{Higgsdoublets}
=1$. In the minimal supersymmetric standard model, one has
\begin{eqnarray}
\left(\begin{array}{c}
 b_1 \\
 b_2 \\
 b_3
\end{array} \right)  =
\left(\begin{array}{c}
 0 \\
 -6 \\
 -9
\end{array} \right) +n_{fam}
\left(\begin{array}{c}
 2 \\
 2 \\
 2
\end{array} \right)+n_{Higgsdoublets}
\left(\begin{array}{c}
 3/10 \\
 1/2 \\
 0
\end{array} \right)
\end{eqnarray}
with $n_{fam}=3$ and $n_{Higgsdoublets}$ = for $\mu > \mu_{threshold}$.
Starting at $\mu = M_Z$ with $\alpha_{\overline{MS}}^{-1}(M_Z) = 127.76
\pm 0.12, \alpha^{-1}_s(M_Z) =0.12\pm 0.01$ and $\sin^2\theta _W^{eff}
\simeq \sin^2\theta _W^{\overline{MS}}(M_Z)=0.2319\pm 0.0007$,
the standard model evolution of the three couplings do not have a
common point, i.e., no grand unification.
But the grand unification is recovered in the minimal supersymmetric
particle threshold at about 1 TeV which is consistent with the lower bound
of the point disappears as soon as $\alpha_s(M_Z) $ becomes greater than
$0.127$. \\
(d) The $Z\rightarrow b\bar{b}$ decay channel provides an attractive
opportunity to study $b-$ particles with reasonable statistics because of
(a longer light time average $\tau_B=1.4 ps$ to be compared to $\tau_{D^{+}}
=1.1ps$ and $\tau_{D^{0}}=0.4 ps $), a bigger cross-section $(\sigma_{B
\bar{B}}\sim 1.2 nb$) and a heaviest mass of about 5 GeV.
The combined LEP measurements
of $\tau_B = 1.40\pm 0.04 ps$ and of the semileptonic branching ratio
$ Br (b\rightarrow l\nu X)=0.112\pm 0.006$ are used to extract the KM matrix
element $|V_{cb}|=0.043\pm 0.005, |V_{ub}/V_{cb}|=0.15\pm 0.10$ and
$\alpha_s(m_b)=0.20\pm 0.03.$
The average preliminary LEP value $\Gamma_b/\Gamma_{had}=0.220\pm 0.0031$
is about $1\sigma $ above the standard model prediction for $m_t > 120 $ GeV
but needs far better accuracy to test the standard model.
As mentioned before, $\Gamma_b$ gets extra contributions from $t\bar{t}W$
and $W^{+}W^{-}t$ triangle graphs within the context of the standard model
and can provide a new test of the standard model through the precise
measurements. But this effect is only about $2\%$ of $\Gamma_b$ for $m_t$
around 150 GeV and poses a challenge to be measured.
The F-B asymmetry $A_{FB}^{b}=\frac{3}{4}A_eA_b=0.098\pm 0.012$
leads to $\sin ^2\theta_W^{eff}=0.2324\pm 0.0020.$
At LEP, as in $p\bar{p}$, the $B^0-\bar{B^0}$ mixing can be studied on the
$b$ quark products. The box diagrams with two $W's$ on two $t's$ in the
intermediate states allow $b\bar{d}\rightarrow \bar{b}d$ and $ b\bar{s}
\rightarrow \bar{b}s $ and induces an oscillation between $B^0-\bar{B^0}$.
The experimental signature of mixing is the observation of like-sign
dileptions from the decays $B^0\rightarrow l^{+}$ and $\bar{B^0}
\rightarrow B^0\rightarrow l^{+}$ and is measured by determining the mixing
parameter $\chi _B=f_d\chi _d+f_s\chi _s$ where $f_d$ and $f_s$ are the
production fraction of $B^0_d$ and $B^0_s$ mesons.
Accurate measurement of the $B^0-\bar{B^0}$ mixing can provide additional test
of the standard model. \\
(e) Of the new particle searches, the lower mass limit for a Higgs boson at
LEP is $m_{H^0}\geq 60 $ GeV while the most recent result from L3 is
$m_{H^0} > 57.7 $ GeV and $m_t > 54 $ GeV from OPAL.
Perhaps the most interesting process observed at LEP is the high mass photon
pair in $e^{+}e^{-}\rightarrow l^{+}l^{-}\gamma \gamma $ events with $M_
{2\gamma }\simeq 60 $ GeV first 4 events reported by L3 and followed by 2
events
by DELPHI.
There are about some 14 events recorded at LEP for $M_{2\gamma } > 50 $ GeV and
now 8 $e^{+}e^{-}, 7 \mu^{+}\mu ^{-}, 3\nu \bar{\nu }$ and $1 q\bar{q}$ events
are reported. L3 calculated the probability $q$ observing 4 or more clustered
events with $M_{2\gamma} > 50 $ GeV is found to be $O(10^{-2})$.
The photon pairs could come from the decay of a massive particle.
While this is certainly a non-standard physics type of events, we [13]
have suggested that the $M_{2\gamma }$ distribution is consistent with a signal
in the muon events as implied by an $"\eta _6"$ production with a mass of
$59 $ GeV, a heavy axion remnant of dynamical electroweak breaking
by a color-sextet quark condensate.
These events pose an exciting possibility to discover the physics beyond  the
standard model.

In short, the recent precision electroweak data on the $Z$ resonance and
parameters are remarkably consistent with the standard model predictions
and no room for new physics seems to be left.
But because of no direct determinations of t-quark and Higgs scalar,
the high order radiative corrections at the best provide a constraining
relation between the t-quark and Higgs masses at the moment.
More precise measurements $M_W$ can narrow down the needed amount of the
radiative corrections and shed light on the better precision tests of the full
electroweak model.
Also, more accurate $t\rightarrow b\bar{b}$ measurements can provide
additional tests of the standard model.
Needless to say the direct discovery of t-quark will be the best test of the
standard electroweak theory.
\section{Acknowledgements}
This work is supported in part by the U.S. Department of Energy, Division
of High Energy Physics contract DE-FG02-91ER40688-Task A.
The author would like to thank Professor H.Song and J.K.Kim and other
colleagues of CTPSNU and KAIST for the warm hospitality
extended to him during his sabbatical leave of absence from Brown University.
He is indebted to the authors of the earlier reviews quoted in Ref.3 and
in particular to Dr. Monica Pepe Altarelli.
Also he wishes to acknowledge S.K.Kang for his assistance in preparing the
manuscript.
\newpage
\section*{ Reference }
\begin{description}
\item[1.] The LEP Collaborations, CERN-PPE/93-157 (1993); see also
          D.Brown in this proceedings.
\item[2.] L.Arnaudon et al, CERN-PPE/92-125 and Phys. Lett. B$\underline
          {284}$, 431 (1992); see also L.Rolandi, XXVI ICHEP 1992,
          CERN-PPE/92-175 (1992).
\item[3.] L.Rolandi (ref.2); J.-E.Augustin, CERN-PPE/93-83 (1993);
          S.C.C.Ting, CERN-PPE/93-34 (1993); M.P.Altarelli, INFN-L.N.F.-
          93/0199(P) (1993). See also P.Langacker, UPR-0555T (1993).
\item[4.] S.L.Glashow, Nucl.Phys. $\underline{22}$, 579 (1961) ;
          S.Weinberg, Phys.Rev.Lett. $\underline{19}$, 1264 (1967) ;
          A.Salam, in Proc.8th Nobel Symp. P.367, ed.N.Svartholm
          (Almqvist and Wiksell, Stockholm, 1968) ; For an introductory
          review of the standard model, see K.Kang, CNRS-IN2P3 Report ISN
92-62.
\item[5.] A.Sirlin, Phys.Rev. D$\underline{22}$, 971(1980) ; D$\underline{29}$,
          89 (1984) ; W.Marciano and A. Sirlin, Phys.Rev.D$\underline{22}$,
          2695 (1980).
\item[6.] W.Hollik, Fortsch.Phys. $\underline{38}$, 165 (1990) ;
          M.Consoli and W.Hollik, and G.Kleiss and F.Jegerlehner in Z Physics
          at LEP 1, G.Altarelli, R.Kleiss and C.Verzegnassi, eds.,
          CERN 89-08 (1989). See also A.Sirlin, Phys.Rev.D$\underline{22}$,
          2695 (1980).
\item[7.] G.'tHooft and M.Veltman, Nucl.Phys.B$\underline{135}$, 365 (1979).
          G.Passarino and M.Veltman, Nucl.Phys.B$\underline{365}$ (1979).
\item[8.] A.Sirlin, Phys.Lett.B$\underline{232}$, 123 (1989) ;
          S.Fanchiatti and A.Sirlin, Phys.Rev.D$\underline{41}$, 319 (1990) ;
          see also, W.Hollik (Ref.6).
\item[9.] F.Berends in Z physics at LEP1 (Ref.6).
\item[10.] G.Burers and F.Jegerlehner (Ref.6).
\item[11.] D.Bardin et al., CERN-TH-6443-92 (1992).
\item[12.] M.Martinez et al., Z.Phys.C$\underline{49}$, 645 (1991)
\item[13.] K.Kang, I.G.Knowles and A.R.White, Mod.Phys.Lett. A$\underline
           {8}$, 1611 (1993).
\end{description}
\end{document}